\documentclass[aps,prd,preprint,nofootinbib]{revtex4}
\usepackage{amsfonts}
\usepackage{mathrsfs}
\usepackage{graphicx}
\usepackage{amsmath}
\usepackage{amssymb}
\usepackage{multirow}
\usepackage{subfigure}
\usepackage{epsfig}
\usepackage{graphicx}
\usepackage{booktabs}
\usepackage{array}
\usepackage{tabularx}
\usepackage{slashed}

\newcommand{\bqa}{\begin{eqnarray}}
\newcommand{\eqa}{\end{eqnarray}}
\newcommand{\beq}{\begin{equation}}
\newcommand{\eeq}{\end{equation}}
\graphicspath{{fig/}{dia/}} \DeclareGraphicsExtensions{.eps}
\begin{document}
\baselineskip 20pt
\title{NLO QCD corrections to pseudoscalar quarkonium production with two heavy flavors in photon-photon collision}
\author{\vspace{1cm} Hao Yang$^1$\footnote[2]{yanghao174@mails.ucas.ac.cn}, Zi-Qiang Chen$^{2}$\footnote[3]{chenziqiang13@mails.ucas.ac.cn}  and Cong-Feng Qiao$^{1,3}$\footnote[1]{qiaocf@ucas.ac.cn, corresponding author} \\}

\affiliation{$^1$ School of Physics, University of Chinese Academy of
Sciences, Yuquan Road 19A, Beijing 100049\\
$^2$ School of Physics and Materials Science, Guangzhou University, Guangzhou 51006, China\\
$^3$ CAS Key Laboratory of Vacuum Physics, Beijing 100049, China\vspace{0.6cm}}

\begin{abstract}

We calculate the next-to-leading order (NLO) quantum chromodynamics (QCD) corrections to $\gamma+\gamma\to \eta_c+c+\bar{c}$, $\gamma+\gamma\to \eta_b+b+\bar{b}$ and $\gamma+\gamma\to B_c+b+\bar{c}$ processes in the framework of non-relativistic QCD (NRQCD) factorization formalism.
The cross sections at the SuperKEKB electron-positron collider, as well as the future collider like the Circular Electron Positron Collider (CEPC), are evaluated.
Numerical results indicate that the NLO corrections are significant, and the uncertainties in theoretical predictions with NLO corrections are reduced as expected.
Due to the high luminosity of the SuperKEKB collider, the $\eta_c+c+\bar{c}$ production is hopefully observable in the near future.

\vspace {5mm} \noindent {PACS number(s): 12.38.Bx, 12.39.Jh, 14.40.Pq, 14.70.Bh}
\end{abstract}

\maketitle

\section{INTRODUCTION}

Heavy quarkonium plays an important role in high energy collider physics, as it presents an ideal laboratory for studying the perturbative and non-perturbative properties of quantum chromodynamics (QCD) within a controlled environment.
The nonrelativistic QCD (NRQCD) factorization formalism \cite{Bodwin:1994jh}, which developed by Bodwin, Braaten, and Lepage, provides a systematic framework for the theoretical study of quarkonium production and decay.
According to the NRQCD factorization formalism, quarkonium production rates can be written as a sum of products of short distance coefficients and the long distance matrix elements (LDMEs).
The short distance coefficients can be calculated as perturbation series in the strong-coupling constant $\alpha_s$, and the LDMEs can be expanded in powers of the relative velocity $v$ of the heavy quarks in the bound state.
In this way, the theoretical prediction takes the form of a double expansion in $\alpha_s$ and $v$.
Although the quarkonium production has been extensively investigated at various colliders, the existing researches are still not sufficient to clarify the underlying production mechanisms \cite{Campbell:2007ws,Gong:2008sn,Ma:2010yw,Butenschoen:2010rq}. 

Quarkonium production in association with two heavy quarks via massless vector boson fusion, i.e. $\gamma(g)+\gamma(g)\to \mathcal{Q}[Q_1\bar{Q_2}]+Q_2+\bar{Q_1}$ process, with $Q_i$ represents $c$ or $b$ quark, is an interesting topic to study.
Experimentally, the techniques to tag heavy quark are now routinely used with high efficiencies, hence the observation of these associated production processes is hopefully possible.
On the other hand, as has been indicated by previous studies \cite{Klasen:2001cu,Qiao:2003ba,Li:2009zzu,Sun:2015hhv,Chen:2016hju,Artoisenet:2007xi,Chang:1992jb,Chang:1994aw,Berezhnoy:1994bb,Kolodziej:1994uu}, these processes are the dominant color-singlet (CS) channels for corresponding single quarkonium inclusive production, and therefore be crucial for pinning down the contributions from CS model.
For $B_c$ meson production, similar mechanism is more important, as quark flavor conservation requires that $B_c$ meson should be produced in accompany with an additional $b\bar{c}$ pair.
Despite the admitted importance, these processes are not fully investigated due to the high technical difficulty.
At present, the only full next-to-leading order (NLO) study is in Ref. \cite{Chen:2016hju}, where the NLO QCD corrections to $\gamma+\gamma\to J/\psi +c+\bar{c}$ process is calculated.
As a further step, in this work, we calculate the NLO QCD corrections to $\gamma+\gamma\to \eta_c +c+\bar{c}$, $\gamma+\gamma\to \eta_b +b+\bar{b}$, and $\gamma+\gamma\to B_c +b+\bar{c}$ processes.

The rest of the paper is organized as follows.
In Sec. II, we present the primary formulae employed in the calculation.
In Sec. III, we elucidate some technical details for the analytical calculation.
In Sec. IV, the numerical evaluation for concerned processes is performed.
The last section is reserved for summary and conclusions.

\section{FORMULATION}
The photon-photon scattering can be achieved at $e^+e^-$ collider like SuperKEKB, where the initial photons are generated by bremsstrahlung effect.
The energy spectrum of bremsstrahlung photon is well formulated by Weizsacker-Williams approximation (WWA) \cite{Frixione:1993yw}:
\begin{equation}
    f_{\gamma}(x) = \frac{\alpha}{2\pi}\left(\frac{1+(1-x)^2}{x}\log\left(\frac{Q^{2}_{\rm max}}{Q^{2}_{\rm min}}\right)+2m^2_{e}x\left(\frac{1}{Q^{2}_{\rm max}}-\frac{1}{Q^{2}_{\rm min}}\right)\right),
\end{equation}
where $Q^{2}_{\rm min}=m^{2}_{e}x^{2}/(1-x)$, $Q^{2}_{\rm max}=(\theta_{c}\sqrt{s}/2)^2(1-x)+Q^{2}_{\rm min}$, $m_e$ is the electron mass, $x=E_{\gamma}/E_{e}$ is the energy fraction of photon, $\sqrt{s}$ is the collision energy of the $e^+e^-$ collider, $\theta_{c}=32$ mrad \cite{Klasen:2001cu} is the maximum scattering angle of the electron or positron.

In future $e^+e^-$ collider like the Circular Electron Positron Collider (CEPC), high energy photon can be achieved through the Compton back-scattering of laser light off electron or positron beam, namely the laser back scattering (LBS) effect.
The LBS photons mostly carry a large energy fraction of the incident electron or positron beam, and at the same time can  achieve high luminosity.
The energy spectrum of LBS photon is \cite{Ginzburg:1981vm}
\begin{equation}
    f_{\gamma}(x)=\frac{1}{N}\left(1-x+\frac{1}{1-x}-4r(1-r)\right),
\end{equation}
where $r=\frac{x}{x_{m}(1-x)}$ and N is the normalization factor:
\begin{equation}
    N=\left(1-\frac{4}{x_{m}}-\frac{8}{x^{2}_{m}}\right)\log(1+x_{m})+\frac{1}{2}+\frac{8}{x_{m}}-\frac{1}{2(1+x_{m})^2}\ .
\end{equation}
Here $x_{m} \approx 4.83$ \cite{Telnov:1989sd}, and the maximum energy fraction of the LBS photon is restricted by $0 \leq x \leq \frac{x_m}{1+x_m}\approx 0.83$.

The total cross section can be obtained by convoluting the $\gamma+\gamma\to \mathcal{Q}[Q_1\bar{Q_2}]+Q_2+\bar{Q_1}$ cross section with the photon distribution functions:
\begin{equation}
    d\sigma=\int dx_{1}dx_{2} f_{\gamma}(x_{1})f_{\gamma}(x_{2})d \hat{\sigma}( \gamma+\gamma\to \mathcal{Q}[Q_1\bar{Q_2}]+Q_2+\bar{Q_1})\ ,
\end{equation}
where $\mathcal{Q} = \eta_c$, $\eta_b$ or $B_c$, and $Q_i$ denotes charm or bottom quark accordingly.
The $d \hat{\sigma}$ is calculated perturbatively up to the NLO level,
\begin{equation}
    d\hat{\sigma}( \gamma+\gamma\to \mathcal{Q}[Q_1\bar{Q_2}]+Q_2+\bar{Q_1})= d\hat{\sigma}_{\rm born} + d\hat{\sigma}_{\rm virtual} + d\hat{\sigma}_{\rm real} + \mathcal{O}(\alpha^{2}\alpha^{4}_{s})\ .
\end{equation}
The Born level cross section, the virtual correction, and the real correction take the following forms:
\begin{equation}
    \begin{split}
        &d\hat{\sigma}_{\rm born}=\frac{1}{2\hat{s}}\overline{\sum}|\mathcal{M}_{\rm tree}|^{2}d{\rm PS}_{3}\ ,\\
        &d\hat{\sigma}_{\rm virtual}=\frac{1}{2\hat{s}}\overline{\sum}2{\rm Re}(\mathcal{M}^{*}_{\rm tree}\mathcal{M}_{\rm oneloop})d{\rm PS}_{3}\ ,\\
        &d\hat{\sigma}_{\rm real}=\frac{1}{2\hat{s}}\overline{\sum}|\mathcal{M}_{\rm real}|^{2}d{\rm PS}_{4}\ ,
    \end{split}
\end{equation}
where $\hat{s}$ is the center-of-mass energy square for the two photons, $\overline{\sum}$ means sum (average) over the polarizations and colors of final (initial) state particles, $d{\rm PS}_{3}$ ($d{\rm PS}_{4}$) denotes final state three (four)-body phase space.

The computation of $d \hat{\sigma}$ can be carried out by using the covariant projection method \cite{Bodwin:2013zu}.
At the leading order of relative velocity expansion, the standard spin and color projection operator can be simplified to
\begin{equation}
       \Pi = \frac{1}{2\sqrt{m_{\mathcal{Q}}}} \gamma^5 (\slashed{p}_{\mathcal{Q}} + m_{\mathcal{Q}})\otimes\left(\frac{\bf{1}_{c}}{\sqrt{N_{c}}}\right),\\
       \label{projection}
\end{equation}
where, $p_{\mathcal{Q}}$ and $m_{\mathcal{Q}}$ are the momentum and mass of the pseudoscalar quarkonium $\mathcal{Q}$ respectively; $\bf{1}_{c}$ represents the unit color matrix,  $N_c=3$ is the number of colors in QCD.

\section{ANALYTICAL CALCULATION}
\begin{figure}[thbp]
    \begin{center}
    \includegraphics[width=0.9\textwidth]{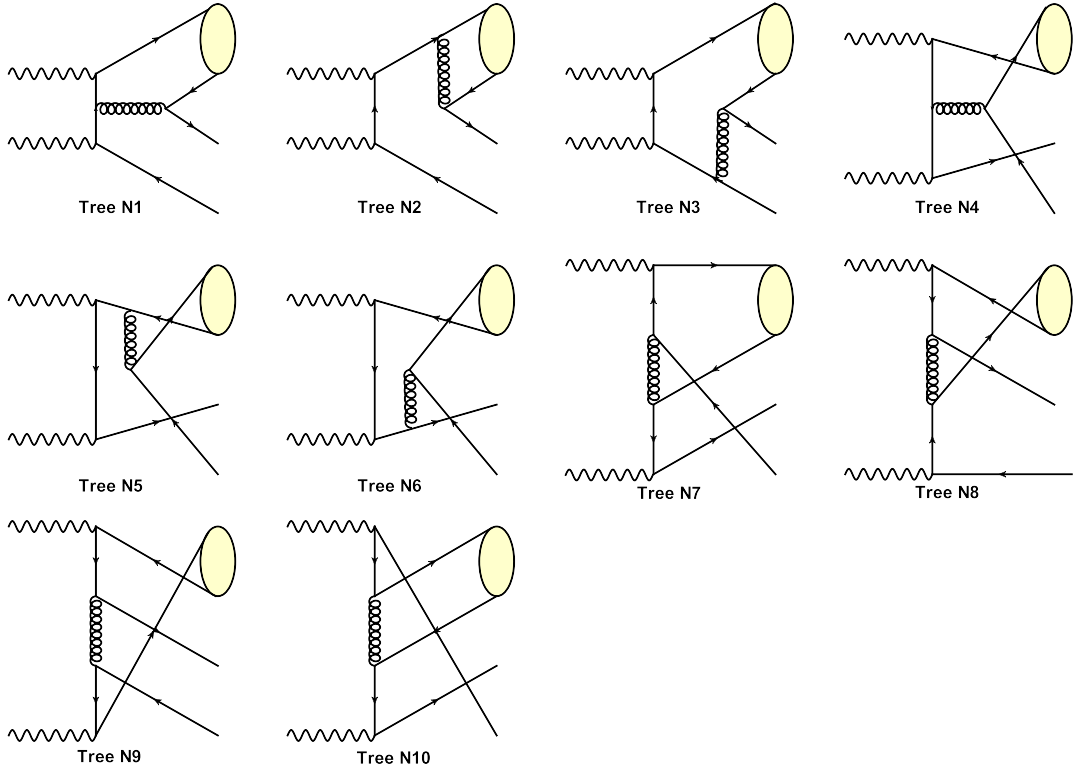}
    \caption{Typical LO Feynman diagrams for $\gamma+\gamma\to \mathcal{Q}[Q_1\bar{Q_2}]+Q_2+\bar{Q_1}$ process.}
    \label{figtree}
    \end{center}
\end{figure}

\begin{figure}[thbp]
    \begin{center}
    \includegraphics[width=0.9\textwidth]{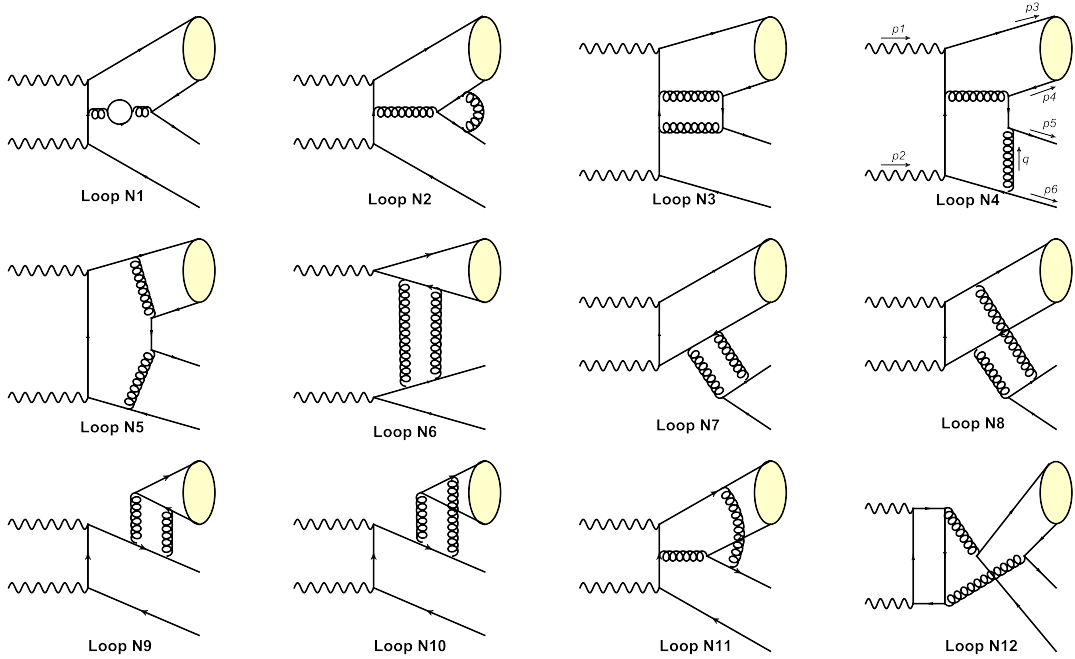}
    \caption{Typical Feynman diagrams in virtual corrections.}
    \label{figloop}
    \end{center}
\end{figure}

At LO, there are twenty Feynman diagrams contributing to the $\gamma+\gamma\to \mathcal{Q}[Q_1\bar{Q_2}]+Q_2+\bar{Q_1}$ process. Half of them are shown in Fig.\ref{figtree}, and the rest can be generated by exchanging the initial two photons.
The typical Feynman diagrams in virtual correction are shown in Fig. \ref{figloop}.
Therein, Loop N1-N5 and Loop N11 arise from the corrections to LO Feynman diagrams, and the rest are new topologies appearing at NLO.
Note, according to the charge-parity conservation, the contributions of type Loop N6 diagrams are vanished, which is verified by our explicit calculation.
And obviously, Loop N6-N10 will not appear in $B_c$ production case.

In the computation of virtual corrections, the conventional dimensional regularization with $D=4-2\epsilon$ is employed to regularize the ultraviolet (UV) divergences, while the infrared (IR) divergences are regularized by introducing an infinitesimal gluon mass $\lambda$.
As a result, the UV and IR singularities appear as $1/\epsilon$ and $\ln (\lambda^2)$ terms, respectively.

In renormalized perturbation theory, the UV singularities are canceled by corresponding counter term diagrams, hence the final virtual corrections are UV finite.
Here, the relevant renormalization constants include $Z_{2}, Z_{3}, Z_{m}$ and $Z_{g}$, which correspond to the heavy quark field, gluon field, heavy quark mass and strong coupling constant, respectively.
Among them, $Z_{2}$ and $Z_{m}$ are defined in the on-mass-shell (OS) scheme, while others are defined in the  modified minimal-subtraction ($\overline{\rm MS}$) scheme. The counterterms are
\begin{equation}
    \begin{split}
        &\delta Z^{\rm OS}_{2}=-C_{F}\frac{\alpha_{s}}{4\pi}\left[2\ln \frac{\lambda^2}{m^2} + \frac{1}{\epsilon_{\rm UV}}-\gamma_{E}+\ln\frac{4\pi\mu^{2}}{m^{2}}+4\right],\\
        &\delta Z^{\rm OS}_{m}=-3C_{F}\frac{\alpha_{s}}{4\pi}\left[\frac{1}{\epsilon_{\rm UV}}-\gamma_{E}+\ln\frac{4\pi\mu^{2}}{m^{2}}+\frac{4}{3}\right],\\
        &\delta Z^{\overline{\rm MS}}_{3}=(\beta_{0}-2C_{A})\frac{\alpha_{s}}{4\pi}\left[\frac{1}{\epsilon_{\rm UV}}-\gamma_{E}+\ln(4\pi)\right],\\
        &\delta Z^{\overline{\rm MS}}_{g}=-\frac{\beta_{0}}{2}\frac{\alpha_{s}}{4\pi}\left[\frac{1}{\epsilon_{\rm UV}}-\gamma_{E}+\ln(4\pi)\right],
        \label{eq_ctterm}
    \end{split}
\end{equation}
where $\gamma_E$ is the Euler's constant; $\mu$ is the renormalization scale, $m$ stands for $m_c$ or $m_b$; $\beta_{0}=\frac{11}{3}C_{A}-\frac{4}{3}T_{F}n_{f}$ is the one-loop coefficient of the QCD $\beta$-function,  $n_{f}$ denotes the active quark flavor numbers; $C_{A}=3, C_{F}=\frac{4}{3}$ and $T_{F}=\frac{1}{2}$ are QCD color factors.
Note, since there is no gluon external leg, the final result is independent of $\delta Z_3$.

The IR singularities in virtual corrections can be isolated by using the method proposed in \cite{Kramer:1995nb}.
Considering the scalar 5-point integral of Fig. \ref{figloop} Loop N4, the IR singularities originate from $q\to 0$ region.
By performing power counting analysis, we have
\begin{equation}
    \begin{split}
       E_0 = 
        &\frac{1}{i\pi^2}\int \frac{d^{4}q}{(2\pi)^4}\frac{1}{q^2-\lambda^2}\frac{1}{(q-p_5)^2-m^2}\frac{1}{(q-p_4-p_5)^2-\lambda^2}\\
       \times&\frac{1}{(q+p_6-p_2)^2-m^2}\frac{1}{(q+p_6)^2-m^2} \\
        \overset{\rm soft}{\sim}&\frac{1}{(p_4+p_5)^2}\frac{1}{(p_6-p_2)^2-m^2} C_0\left(p_6^2,(p_5+p_6)^2,p_5^2,\lambda^2,m^2,m^2\right)\\
        \overset{\rm soft}{\sim}&\frac{1}{s_{45}(t_{26}-m^2)}\frac{\ln(\lambda^2)}{\sqrt{s_{56}(s_{56}-4m^2)}}\left(\ln\frac{\sqrt{s_{56}}-\sqrt{s_{56}-4m^2}}{\sqrt{s_{56}}+\sqrt{s_{56}-4m^2}}+i\pi\right),
    \end{split}
\end{equation}
where $s_{45}=(p_4+p_5)^2$, $s_{56}=(p_5+p_6)^2$, $t_{26}=(p_2-p_6)^2$.

The Coulomb singularities, which appears when a potential gluon is exchanged between the constituent quarks of a quarkonium, are also regularized by infinitesimal gluon mass $\lambda$.
We obtain
\begin{equation}
    2{\rm Re}(\mathcal{M}^{*}_{\rm tree}\mathcal{M}_{\rm oneloop}) \overset{\rm Coulomb}{\sim} |\mathcal{M}_{\rm tree}|^{2} \frac{2\alpha_s C_F m}{\lambda},
    \label{eq_coulomb}
\end{equation}
which can be absorbed into the wave function of quarkonium.
Note, for $B_c$ production, the $m$ in Eq. (\ref{eq_coulomb}) should be replaced by $\frac{2m_bm_c}{m_b+m_c}$.

\begin{figure}[thbp]
    \begin{center}
    \includegraphics[width=0.9\textwidth]{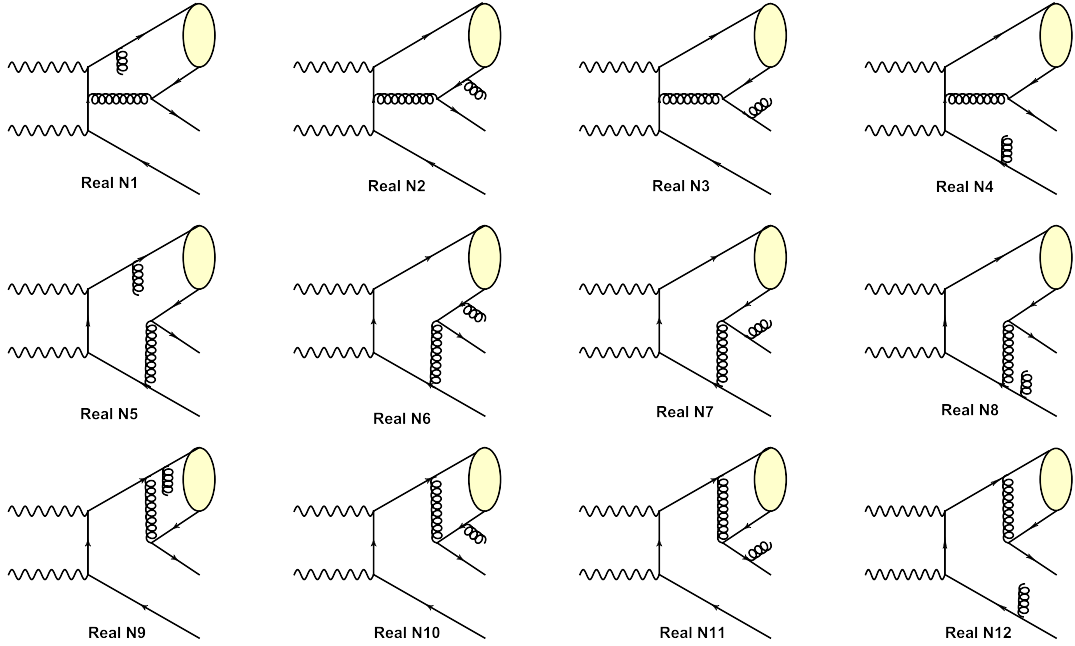}
    \caption{Typical Feynman diagrams in real corrections.}
    \label{figreal}
    \end{center}
\end{figure}

Typical Feynman diagrams in real corrections are shown in Fig. \ref{figreal}.
The IR divergences here are also regularized by infinitesimal gluon mass.
To isolate the IR singularities, the subtraction method which formulated in Ref. \cite{Dittmaier:1999mb} is employed.
As a result, the contribution of real corrections can be separated into two parts:
\begin{equation}
    \hat{\sigma}_{\rm real}=\hat{\sigma}_{\rm real}^A + \hat{\sigma}_{\rm real}^B, 
\end{equation}
with
\begin{align}
    \hat{\sigma}_{\rm real}^A &= \frac{1}{2\hat{s}}\int d{\rm PS}_{4} (\overline{\sum}|\mathcal{M}_{\rm real}|^{2} - |\mathcal{M}_{\rm sub}|^{2}), \\
    \hat{\sigma}_{\rm real}^B &= \frac{1}{2\hat{s}}\int d{\rm PS}_{4} |\mathcal{M}_{\rm sub}|^{2} = \frac{1}{2\hat{s}}\int d{\rm PS}_{3} \int [dp_g]|\mathcal{M}_{\rm sub}|^{2}.
\end{align}
Here $[dp_g]$ denotes the phase space of the additional emitted gluon, $|\mathcal{M}_{\rm sub}|^{2}$ is an auxiliary subtraction function which holds the same asymptotic behavior as $\overline{\sum}|\mathcal{M}_{\rm real}|^{2}$ in the soft limits. 
Hence the difference $(\overline{\sum}|\mathcal{M}_{\rm real}|^{2} - |\mathcal{M}_{\rm sub}|^{2})$ is non-singular at each point of phase space, and the integral can be evaluated with $\lambda=0$ everywhere.
With an appropriate construction of $|\mathcal{M}_{\rm sub}|^{2}$ \cite{Dittmaier:1999mb}, the integral $ \int [dp_g]|\mathcal{M}_{\rm sub}|^{2}$ can be carried out analytically.
After adding $2{\rm Re}(\mathcal{M}^{*}_{\rm tree}\mathcal{M}_{\rm oneloop})$ and $ \int [dp_g]|\mathcal{M}_{\rm sub}|^{2}$, the IR singularities, i.e. $\ln(\lambda^2)$ terms, cancel with each other as expected.

\section{NUMERICAL RESULTS}
In the numerical calculation, the input parameters are taken as
\begin{align}
&\alpha=1/137.065,\quad m_e=0.511\ {\rm MeV},\quad m_c=1.5\ {\rm GeV},\quad m_b=4.8\ {\rm GeV};\nonumber \\
&|R_{\eta_c}^{\rm LO}(0)|^2=0.528\ {\rm GeV}^3,\quad |R_{\eta_c}^{\rm NLO}(0)|^2=0.907\ {\rm GeV}^3; \nonumber \\
&|R_{\eta_b}^{\rm LO}(0)|^2=5.22\ {\rm GeV}^3,\quad |R_{\eta_b}^{\rm NLO}(0)|^2=7.48\ {\rm GeV}^3; \nonumber \\
&|R_{B_c}(0)|^2=1.642\ {\rm GeV}^3 \nonumber.
\end{align}
Here, the $B_c$ wave function at the origin is estimated by using the Buchmueller-Tye potential \cite{Eichten:1994gt}.
According to the heavy quark spin symmetry of NRQCD at the leading order in relative velocity expansion \cite{Bodwin:1994jh}, here we take $R_{\eta_c}(0)=R_{J/\psi}(0)$ and $R_{\eta_b}(0)=R_{\Upsilon}(0)$.
The $J/\psi$ and $\Upsilon$ radial wave functions at the origin are extracted from their leptonic widths.
\begin{equation}
    \Gamma(\mathcal{Q}\to e^+e^-)=\frac{\alpha^2e_Q^2}{m_Q^2}|R_{\mathcal{Q}}(0)|^2\left(1-4C_F\frac{\alpha_s(\mu_0)}{\pi}\right),\ e^{}_Q=\begin{cases}\frac{2}{3},\ {\rm if}\ \mathcal{Q}=J/\psi\\ \frac{1}{3},\ {\rm if}\ \mathcal{Q}=\Upsilon\end{cases},
\end{equation}
with $\mu_0=2m^{}_Q$, $\Gamma(J/\psi\to e^+e^-)=5.55$ keV, and $\Gamma(\Upsilon\to e^+e^-)=1.34$ keV \cite{ParticleDataGroup:2020ssz}.
Note, the LO and NLO extractions are employed in the LO and NLO calculation respectively.

In the NLO calculation, the two-loop formula
\begin{equation}
    \frac{\alpha_{s}(\mu)}{4\pi}=\frac{1}{\beta_{0}L}-\frac{\beta_{1}\ln L}{\beta^{3}_{0}L^{2}},
\end{equation}
for the running coupling constant is employed, in which, $L=\ln (\mu^{2}/\Lambda^{2}_{\rm QCD})$, $\beta_0=\tfrac{11}{3}C_A-\tfrac{4}{3}T_Fn_f$, $\beta_{1}=\frac{34}{3}C^{2}_{A}-4C_{F}T_{F}n_{f}-\frac{20}{3}C_{A}T_{F}n_{f}$, with $n_f=4$, $\Lambda_{\rm QCD}=297$ MeV for $\eta_c$ production, and $n_f=5$, $\Lambda_{\rm QCD}=214$ MeV for $\eta_b$ and $B_c$ production.
For LO calculation, the one-loop formula for the running coupling constant is used.

\begin{table}[ht]
    \caption{The LO and NLO total cross sections for $\eta_c+c+\bar{c}$ production via photon-photon fusion at the SuperKEKB collider. Here $m_c=1.5$ GeV, $\mu=r\sqrt{4m_c^2 + p_{t}^2}$ with $r=\{0.5,1,2\}$. The transverse momentum cut $ 0.2\ {\rm GeV} \le p_{t} \le 4.0$ GeV is imposed to $\eta_c$ meson.}
    \begin{center}
       \begin{tabular}{p{2.5cm}<{\centering} p{2.5cm}<{\centering} p{2.5cm}<{\centering} p{2.5cm}<{\centering}}
        \toprule
        \hline
            $r$  & $0.5$  &  $1$ & $2$ \\
        \hline
        $\sigma_{\rm LO}$ (fb)&
        $0.340$ &
        $0.171$ &
        $0.102$ \\
        $\sigma_{\rm NLO}$ (fb)&
        $0.647$ &
        $0.366$ &
        $0.239$ \\
        \botrule
      \end{tabular}
    \end{center}
    \label{tabscale}
\end{table}

\begin{table}[ht]
    \caption{The LO and NLO total cross sections for $\eta_c+c+\bar{c}$ production via photon-photon fusion at the SuperKEKB collider. Here $m_c=\{1.4,1.5,1.6\}$ GeV, $\mu=\sqrt{4m_c^2 + p_{t}^2}$. The transverse momentum cut $ 0.2\ {\rm GeV} \le p_{t} \le 4.0$ GeV is imposed to $\eta_c$ meson.}
    \begin{center}
       \begin{tabular}{p{2.5cm}<{\centering} p{2.5cm}<{\centering} p{2.5cm}<{\centering} p{2.5cm}<{\centering}}
        \toprule
        \hline
            $m_c\ (\rm{GeV})$  & $1.4$  &  $1.5$ & $1.6$ \\
        \hline
        $\sigma_{\rm LO}$ (fb)&
        $0.393$ &
        $0.171$ &
        $0.074$ \\
        $\sigma_{\rm NLO}$ (fb)&
        $0.821$ &
        $0.366$ &
        $0.161$ \\
        \botrule
      \end{tabular}
    \end{center}
    \label{tabmc}
\end{table}

We investigate the production of $\eta_c+c+\bar{c}$ with WWA photons as the initial state at the SuperKEKB collider,  where the beam energies of the positron and electron are 4 GeV and 7 GeV respectively, yielding a center-of-mass energy of 10.6 GeV. 
In order to estimate the theoretical uncertainties induced by renormalization scale and charm quark mass, we set $\mu=r\sqrt{4m_c^2+p_t^2}$ with $r=\{0.5,1,2\}$, and $m_c=\{1.4,1.5,1.6\}$ GeV.
The corresponding results are shown in Table \ref{tabscale} and Table \ref{tabmc}, respectively.
It can be seen that the NLO corrections are significant, and the total cross sections are enhanced by a factor (defined as the $K$ factor) of about $2.1$.
To measure the dependency of the cross section on renormalization scale and charm quark mass, we define $R_{\mu}=\frac{\sigma|_{r=0.5}-\sigma|_{r=2}}{\sigma|_{r=1}}$ and $R_{m}=\frac{\sigma|_{m_c=1.4}-\sigma|_{m_c=1.6}}{\sigma|_{m_c=1.5}}$.
Then we have $R_\mu^{\rm LO}=1.39$, $R_\mu^{\rm NLO}=1.11$, $R_{m}^{\rm LO}=1.86$ and $R_{m}^{\rm NLO}=1.80$, which indicates that the theoretical uncertainties are slightly reduced by NLO corrections.

In the coming year, the instantaneous luminosity of the SuperKEKB collider may reach $8\times10^{35}\ {\rm cm}^{-2}{\rm s}^{-1}$ \cite{urlsuperkekb}.
Then the yearly produced $\eta_c+ c+\bar{c}$ events is estimated to be $(4.05\sim 20.7)\times 10^3$.
In experiment, $\eta_c$ can be reconstructed through its $K\bar{K}\pi$ decay channel with the branching ratio ${\rm Br}(\eta_c\to K\bar{K}\pi)=7.3\%$ \cite{ParticleDataGroup:2020ssz}, and the tagging efficiency of charm quark is about $41\%$ \cite{ATLAS:2018mgv}.
Therefore we expect to obtain $49\sim 253$ $\eta_c+ c+\bar{c}$ events per year.

\begin{table}[ht]
    \caption{The LO (in brackets) and NLO total cross sections for $\eta_c +c+\bar{c}$, $\eta_b+ b+\bar{b}$, $B_c+ b+\bar{c}$ production via photon-photon fusion at $\sqrt{s}=250$ GeV. Here the cut 1 GeV$\le p_{t} \le 50$ GeV is imposed.}
    \begin{center}
        \begin{tabular}{p{2.5cm}<{\raggedright} p{2.5cm}<{\raggedright} p{2.5cm}<{\raggedright} p{2.5cm}<{\raggedright}}
        \toprule
        \hline
            photon  & $\sigma_{\eta_c c\bar{c}}$ (fb)  &  $\sigma_{\eta_b b\bar{b}}$ (fb) & $\sigma_{B_c b\bar{c}}$ (fb) \\
        \hline
        WWA&
        $217.8(126.7)$ &
        $0.073(0.055)$ &
        $1.051(0.778)$ \\

        LBS&
        $1127(606)$ &
        $3.84(2.08)$ &
        $34.0(23.1)$ \\
        \botrule
      \end{tabular}
    \end{center}
    \label{tabcepc}
\end{table}

Of the future high energy $e^+e^-$ colliders, like the CEPC, the collision energy may reach 250 GeV \cite{CEPCStudyGroup:2018ghi}.
And the LBS photon collision can be realized by imposing a laser beam to each $e$ beam.
Therefore, we investigate the $\eta_c +c+\bar{c}, \eta_b+ b+\bar{b}$ and $B_c+b+\bar{c}$ productions under both WWA and LBS photon collisions with $\sqrt{s}=250$ GeV.
The corresponding LO and NLO total cross sections are presented in Table.\ref{tabcepc}.
As the energy scale of CEPC is higher than that of SuperKEKB, the $K$ factors here are less than 2.
Taking a typical luminosity $\mathcal{L}=10^{34}\ {\rm cm}^{-2}{\rm s}^{-1}$ \cite{ParticleDataGroup:2020ssz}, the number of reconstructed $\eta_c+ c+\bar{c}$ candidates per year is about  $1.15 \times 10^4$ for the WWA photon case, $5.97 \times 10^4$ for the LBS photon case.
For $\eta_b+ b+\bar{b}$ production, the observation is somewhat difficult due to the insignificant production rates.
For $B_c$ inclusive production, it is not necessary to reconstruct the produced $b$ and $\bar{c}$ jets.
Assuming $B_c$ is reconstructed through $B_c^{\pm} \rightarrow J/\psi(1S)\pi^{\pm}$, whose branching fraction is predicted to be $0.5\%$ \cite{Chang:1992pt}, and $J/\psi$ is reconstructed through $J/\psi \rightarrow l^+l^-(l=e,\mu)$ with a branching fraction of about $12\%$ \cite{ParticleDataGroup:2020ssz}, the number of the reconstructed $B_c$ candidates for LBS photon case would reach 12 per year.
Note, since $B_c^*$ almost all decays to $B_c$, a more exact prediction on $B_c$ candidates should take into account the $\gamma+\gamma\to B_c^*+b+\bar{c}$ process, and we leave it for future study.

\begin{figure}[!thbp]
    \centering
    \subfigure[]{\includegraphics[width=0.49\textwidth]{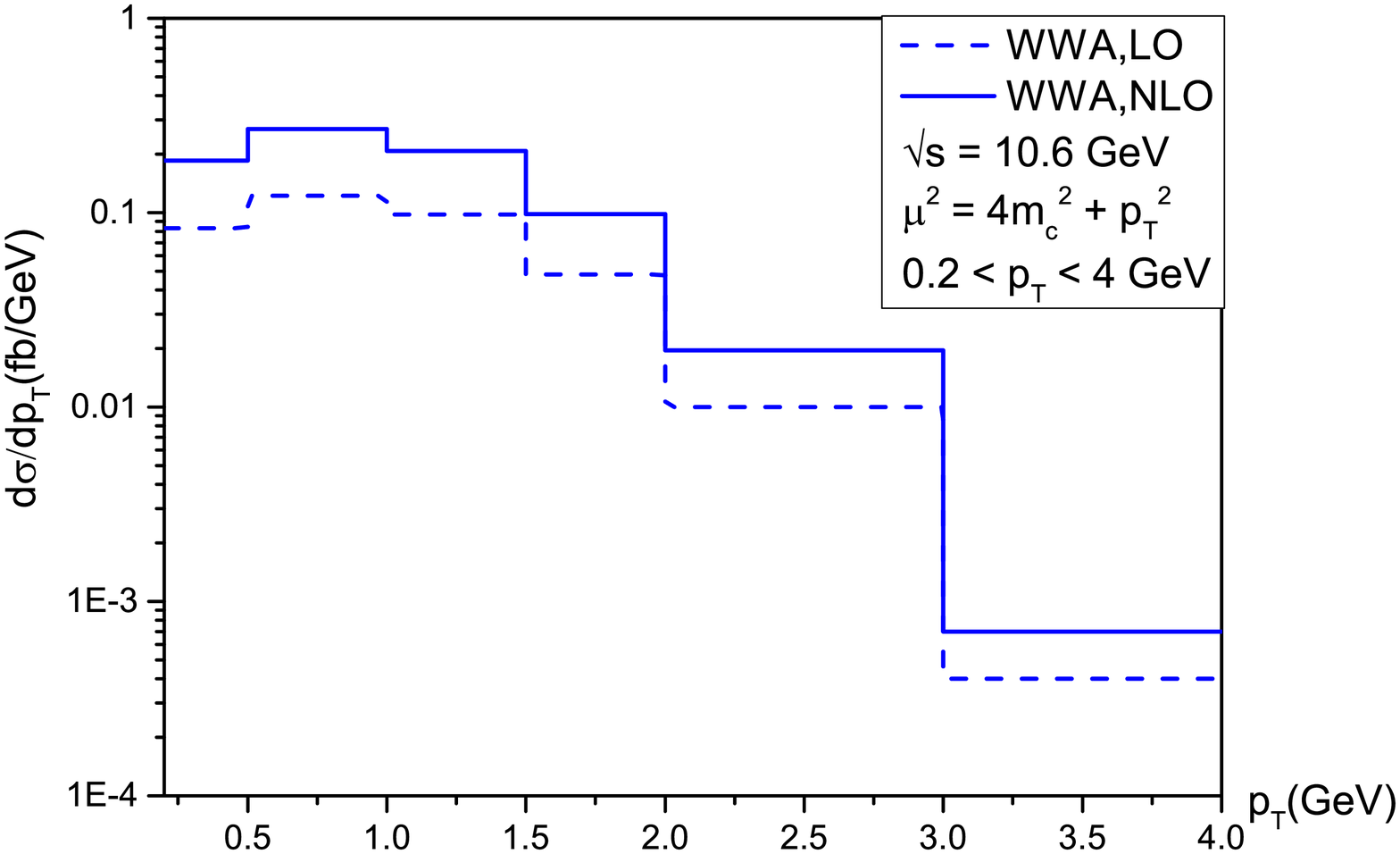}}
    \subfigure[]{\includegraphics[width=0.49\textwidth]{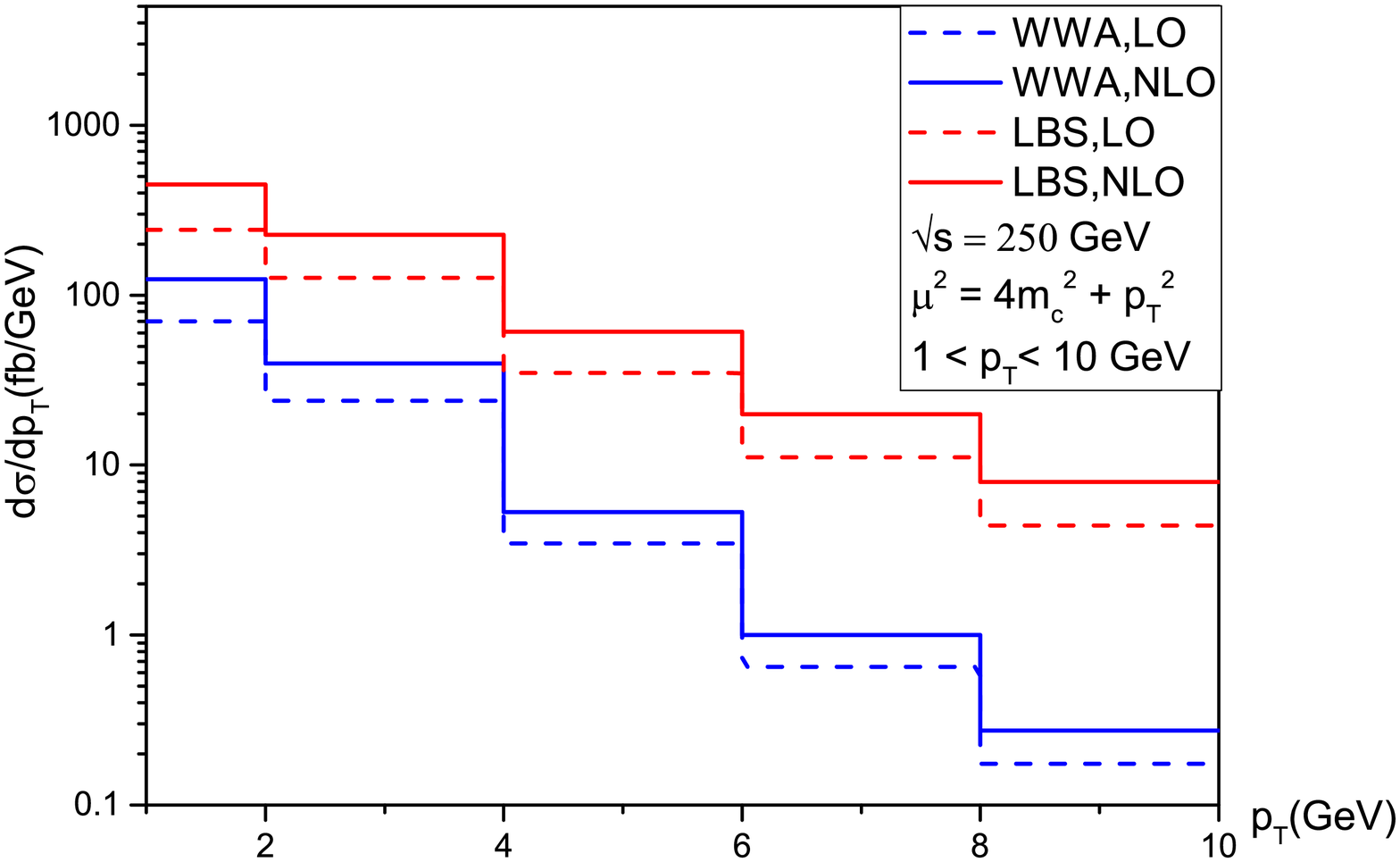}}
    \caption{The $p_t$ distribution for the $\eta_c c\bar{c}$ production via photon-photon fusion at (a) the SuperKEKB collider; (b) the CEPC. Here the renormalization scale $\mu=\sqrt{4m_{c}^2+p^{2}_{t}}$, the transverse momentum cut $0.2 \le p_{t} \le 4$ GeV and $1 \le p_{t} \le 10$ GeV is imposed on $\eta_c$ respectively.}
    \label{figpt}
\end{figure}

As the number of events corresponding to $\eta_c+c+\bar{c}$ production is large, it is worthy to perform a more elaborate phenomenological analysis.
The differential cross sections versus $p_t$, i.e. the transverse momentum of $\eta_c$, at the SuperKEKB collider and CEPC are shown in Fig.\ref{figpt}. 
It can be seen that the NLO corrections cause an upward shift of the LO distributions, and leave the shapes nearly unchanged.

\section{SUMMARY AND CONCLUSIONS}
In this work, we investigate the $\eta_c+c+\bar{c}$, $\eta_b+b+\bar{b}$, $B_c+b+\bar{c}$ production via photon-photon fusion at the NLO QCD accuracy in the framework of NRQCD factorization formalism.
The total cross sections and the differential cross sections versus transverse momentum at the SuperKEKB collider and the CEPC are given.

Numerical results shows that, after including the NLO corrections, the total cross sections are significantly enhanced, and their dependences on renormalization scale and heavy quark mass parameter are reduced as expected.
Due to the high luminosity of the SuperKEKB collider, the $\eta_c +c+\bar{c}$ production via photon-photon fusion is hopeful to be observed in the near future.
At the higher energy collider like CEPC, the production rate of $\eta_c +c+\bar{c}$ is largely enhanced, which leads to inspiring events number.
If the LBS photon collision can be realized, the observation of $B_c$ meson production at the CEPC can also be expected.

\vspace{1.4cm} {\bf Acknowledgments}
This work was supported in part by the National Key Research and Development Program of China under Contracts No. 2020YFA0406400,
and the National Natural Science Foundation of China (NSFC) under the Grants No. 11975236, No. 11635009, and  No. 12047553.


\end{document}